\documentclass[aps,floats,twocolumn,epsf,prl,showpacs]{revtex4-1}
\usepackage{graphicx}
\usepackage{epstopdf}
\usepackage{amsmath}
\usepackage{amssymb}
\usepackage[colorlinks=true,urlcolor=blue,citecolor=blue,linkcolor=blue,breaklinks=true]{hyperref}
\usepackage{color}
\usepackage[dvipsnames]{xcolor}

\usepackage{orcidlink}
\newcommand{\lsc}[1]{}
\newcommand{\khr}[1]{{\color{green}}}

\begin{document}

\title{Electronic structure of the putative room-temperature superconductor
  Pb$_9$Cu(PO$_4$)$_6$O}

\author{Liang Si$^{a,b}$\orcidlink{0000-0003-4709-6882} and Karsten Held$^b$\orcidlink{0000-0001-5984-8549}}
\affiliation{$^a$ School of Physics, Northwest University, Xi'an 710127, China\\$^b$ Institute of Solid State Physics, TU Wien, 1040 Vienna, Austria}
  
\email{liang.si@ifp.tuwien.ac.at}
\email{held@ifp.tuwien.ac.at}

\date{\today}

\begin{abstract}
A recent paper [Lee {\em et al.}, J. Korean Cryt. Growth Cryst. Techn. {\bf 33}, 61 (2023)] provides some experimental indications that  Pb$_{10-x}$Cu$_x$(PO$_4$)$_6$O with $x\approx 1$, coined LK-99, might be a room-temperature superconductor  at ambient pressure.
Our density-functional theory calculations show lattice parameters and a volume contraction with $x$ -- very similar to experiment.  The DFT electronic structure shows Cu$^{2+}$ in a $3d^9$ configuration with two flat Cu bands  crossing the Fermi energy.
This puts  Pb$_{9}$Cu(PO$_4$)$_6$O in an ultra-correlated regime and suggests that, without doping, it is a Mott or charge transfer insulator.  If doped such an electronic structure might support flat-band superconductivity or  an correlation-enhanced electron-phonon mechanism, whereas a diamagnet without superconductivity appears to be rather at odds with our results.
  \end{abstract}

\maketitle

\noindent
Hitherto, milestones of superconductivity research  include its discovery by Onnes in 1911, the BCS theory of Bardeen, Cooper and Schrieffer~\cite{BCS1957},
the discovery of high--temperature superconductors by Bednorz and M\"uller~\cite{Bednorz1986}, and, more recently that of hydride superconductors by Eremets and coworkers~\cite{Eremets_Nature_2015_SH3} as well as nickelate superconductivity by Li {\em et al.}~\cite{li2019superconductivity,zeng2020,Li2020dome}. Ever since Onnes, finding a room temperature superconductor has been {\em the} big dream of condensed matter physics. Such a superconductor would revolutionize
the way we generate, transport and consume electric energy.

\begin{figure}[tb]
\includegraphics[width=\linewidth,angle=0]{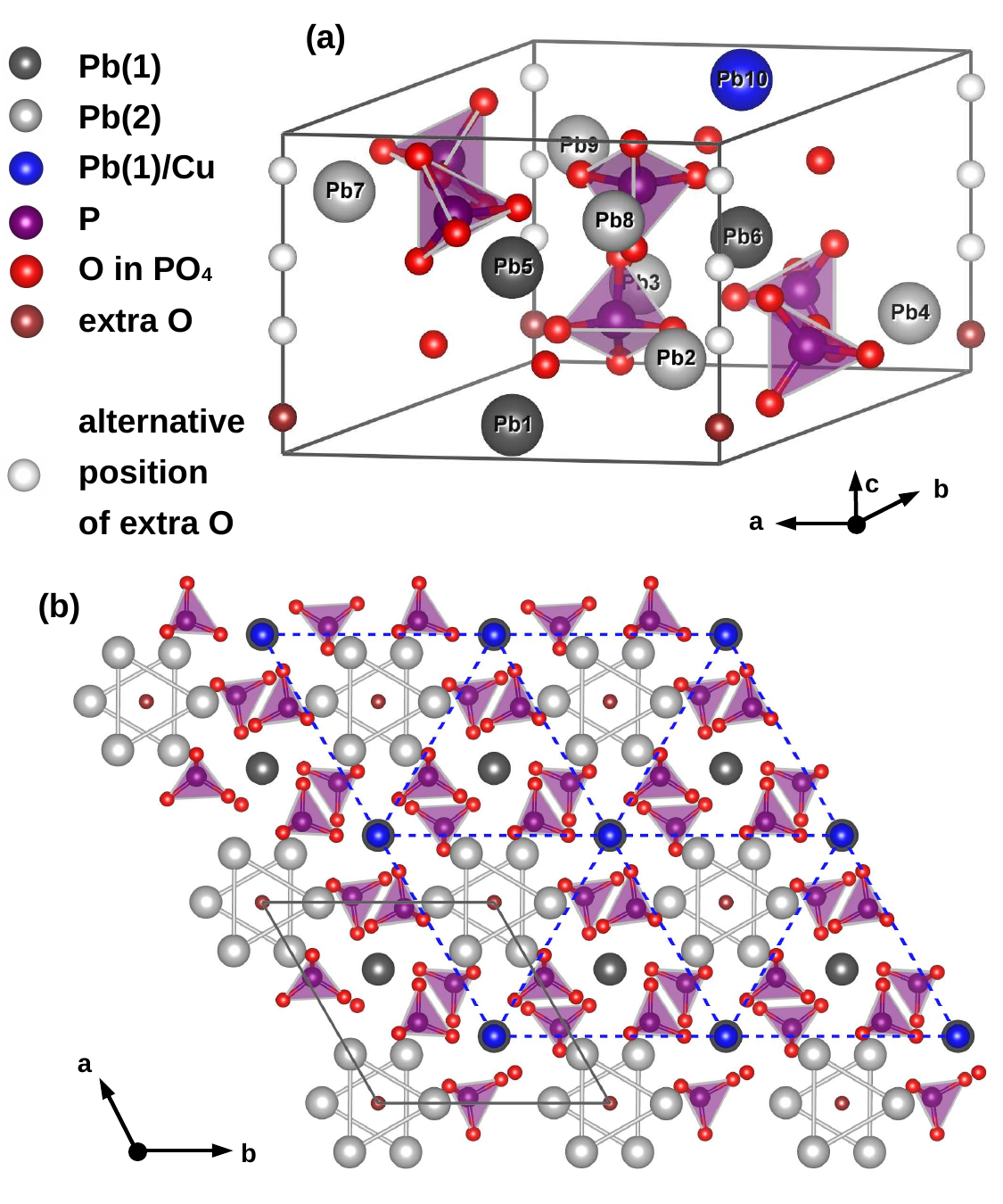}
        \caption{Crystal structure of Pb$_{9}$Cu(PO$_4$)$_6$O: (a) single unit cell in side view; (b) four unit cells plus surrounding atoms  in top view.  There is some uncertainty regarding the position of the one extra O per formula unit and at which Pb site Cu is substituted. The extra O can occupy the dark red position as indicated in (a) or one of the empty white circles instead. As for the 10 Pb atoms in the unit cell, these fall into two symmetry classes: Pb(1) [dark gray] and Pb(2) [light gray]. The latter are located around the extra O, see (b). Our DFT calculations show that it is however energetically favorable for the Cu atom to be as far away from the extra O as possible, i.e., to occupy the blue Pb10 position of the Pb(2) class. This  leaves us --for a lead-apatite structure and a  single formula unit as unit cell-- with the structure displayed.
        \label{Fig:CrystStruct}}
\end{figure}

Lee {\em et al.}~\cite{Lee2023_1,Lee2023_2,Lee2023_3}  proclaim that they have successfully synthesized such a room temperature superconductor, Pb$_{10-x}$Cu$_x$(PO$_4$)$_6$O with $0.9<x<1$, even at ambient pressure.
This is indicated by (i) a drastic drop in the resistivity~\cite{Lee2023_1,Lee2023_3} (according to~\cite{Lee2023_2} to the order of $10^{-10} - 10^{-11}\,\Omega$cm despite a quite substantial noise level~\cite{Lee2023_1,Lee2023_3}); (ii) a negative (diamagnetic) susceptibility and levitation of the superconductor on a magnet~\cite{Lee2023_3};  (iii)  
extraordinarily sharp voltage jumps at the critical currents~\cite{Lee2023_1,Lee2023_2,Lee2023_2} with 
a vanishing critical current strength at temperatures of about 400$\,$K and fields of about 3000$\,$Oe~\cite{Lee2023_2}. The recipe to synthesize   Pb$_{10-x}$Cu$_x$(PO$_4$)$_6$O appears easy enough~\cite{Lee2023_3} for other groups  to follow suit. Thus further experiments will reveal whether Pb$_{9}$Cu(PO$_4$)$_6$O is indeed the first room-temperature superconductor or not.

In any case, the experiments by Lee {\em et al.}~\cite{Lee2023_1,Lee2023_2,Lee2023_3} are exiting and definitely call for a more thorough theoretical understanding of this rather unusual material. The crystal structure of Pb$_{9}$Cu(PO$_4$)$_6$O as obtained by x-ray diffraction (XRD)~\cite{Lee2023_1,Lee2023_2,Lee2023_3} is a modified lead-apatite structure and is shown in Fig.~\ref{Fig:CrystStruct}. The first step to obtain a theoretical understanding of a new material is an impartial density-functional theory (DFT) calculation.
 
In this paper, we perform such DFT calculations, focusing on the crystal and electronic structure. The latter can serve at least as a starting point for subsequent many-body calculations.
Relaxing the lead-apatite crystal structure, we find very similar lattice parameters as Lee {\em et al.}~\cite{Lee2023_1,Lee2023_3}. The calculation further confirms the observed lattice compression  when substituting Pb by Cu. The DFT electronic structure of the parent compound ($x=0$) is insulating, whereas there are two very flat predominately Cu $d$-bands crossing the Fermi energy for $x=1$. These flat bands are prone to be split into Hubbard bands. In this case undoped Pb$_{9}$Cu(PO$_4$)$_6$O would become a Mott or charge transfer insulator, depending on the relative position of the lower Hubbard band and the other bands.  

{\sl Computational method}---
For the DFT structural relaxations and electronic structure calculations, we use \textsc{Vasp}~\cite{PhysRevB.47.558,kresse1996efficiency} (projected augmented plane waves) with the GGA-PBESol~\cite{PhysRevLett.100.136406} exchange-correlation potential. More details are available in the Supplemental Material Section~I \cite{SM}.

{\sl Crystal structure}---
Let us start with the undoped parent compound  Pb$_{10}$(PO$_4$)$_6$O, a lead apatite with a hexagonal structure (P6$_3$/m, 176) identical to that displayed in Fig.~\ref{Fig:CrystStruct} \cite{krivovichev2003crystal}. 
For the parent compound, there is still the uncertainty at which position (out of four) the one oxygen is to be placed: open circles or dark red circle at the edge(s) of Fig.~\ref{Fig:CrystStruct}~(a).
Here, we refer to this O as the ``extra O", as it is not part of the six PO$_4$ tetrahedra.
For a single unit cell all of these O positions are however equivalent. We have relaxed the structure in DFT and find lattice parameters in Table~\ref{Tab1_lattice} to agree with experiment. 
We have further checked all possible  distributions of four Os in the larger supercell of Fig.~\ref{Fig:CrystStruct} (b) consisting of four formula units, see Supplemental Material~\cite{SM} Section~I. The energy difference between the different oxygen arrangements is only $\sim$6$\,$meV per unit cell, which corresponds to 70$\,$K and is hence not relevant at room temperature.

\begin{table}[tbh]
\begin{tabular}{c|c|c|c|c}
\hline
\hline
Phase   & $a$(\AA) & $c$(\AA) & $V$(\AA$^3$) & From \\
\hline
Pb$_{10}$(PO$_4$)$_6$O    & 9.865 & 7.431 & 626.28 & exp.~\cite{Lee2023_1,Lee2023_3}  \\
\hline
Pb$_{10}$(PO$_4$)$_6$O    & 9.825 & 7.371 & 616.22 & this work (GGA-PS) \\
\hline
\hline
Pb$_9$Cu(PO$_4$)$_6$O   & 9.843 & 7.428 & 623.24 & exp.~\cite{Lee2023_1,Lee2023_3}  \\
\hline
Pb$_9$Cu(PO$_4$)$_6$O    & 9.661 & 7.226 & 584.04 & this work (GGA-PS) \\
\hline
\hline
\end{tabular}
\caption{DFT relaxed lattice parameters and unit cell volume compared to experiment.
\label{Tab1_lattice}}
\end{table}

Next we study the crystal structure of the putative superconductor Pb$_{9}$Cu(PO$_4$)$_6$O. 
The XRD data of {\em Lee et al.} \cite{Lee2023_1,Lee2023_2,Lee2023_3} show a modified lead-apatite structure. In one unit cell of this structure there are 4$\times$10 different arrangement of the extra O and doped Cu, some of them related by symmetry. We have calculated all possibilities, see Supplemental Material~\cite{SM} Section~I, and find that Cu prefers to occupy the position that is farthest away from O \cite{Lee2023_2,Lee2023_3}. This results in the supercell displayed in Fig.~\ref{Fig:CrystStruct} (a) with Cu occupying one of the four dark gray/blue  Pb(1) sites. The energy gain compared to other Cu-O arrangement is at least 12.1\,meV, see Supplemental Material~\cite{SM} Section I~B.

Indeed it was already discussed in \cite{Lee2023_1,Lee2023_2,Lee2023_3} that the six Pb sites around the tube where the extra O is located [light gray Pb(2) sites in  Fig.~\ref{Fig:CrystStruct}] are not substituted with Cu, that Cu instead occupies one of the dark gray  sites [Pd(1)] further away from the extra O in agreement with our DFT calculation.
Our DFT crystal structure also confirms the volume reduction compared to the parent compound, see Table~\ref{Tab1_lattice}, albeit it is considerably larger in DFT than in experiment.

The periodic continuation of the single unit cell is shown in the the top view of  Fig.~\ref{Fig:CrystStruct} (b). Here, the Cu atoms (blue) arrange in a two-dimensional triangular lattice and the Pb(1) atoms (dark gray)
in a similar triangular lattice in essentially the same layer, see Fig.~\ref{Fig:CrystStruct}~(b). In the $c$
direction, this is interlaced by a similar layer of only Pb(1) atoms that sit exactly beneath the first layer and are thus not visible in Fig.~\ref{Fig:CrystStruct}~(b).
The Pb(2) atoms (light gray) arrange instead in a hexagon
(two triangles in different layers) around the position (channel) of the  extra O.
We cannot exclude a more complex long-range arrangement of the Cu and extra O sites based on a larger unit cell at the moment. While this can change the lattice of the Cu sites substantially, the main findings presented in the {\sl Conclusion} are not affected.

\begin{figure*}[tb]
\begin{minipage}{0.685\textwidth} 
\includegraphics[width=\linewidth,angle=0]{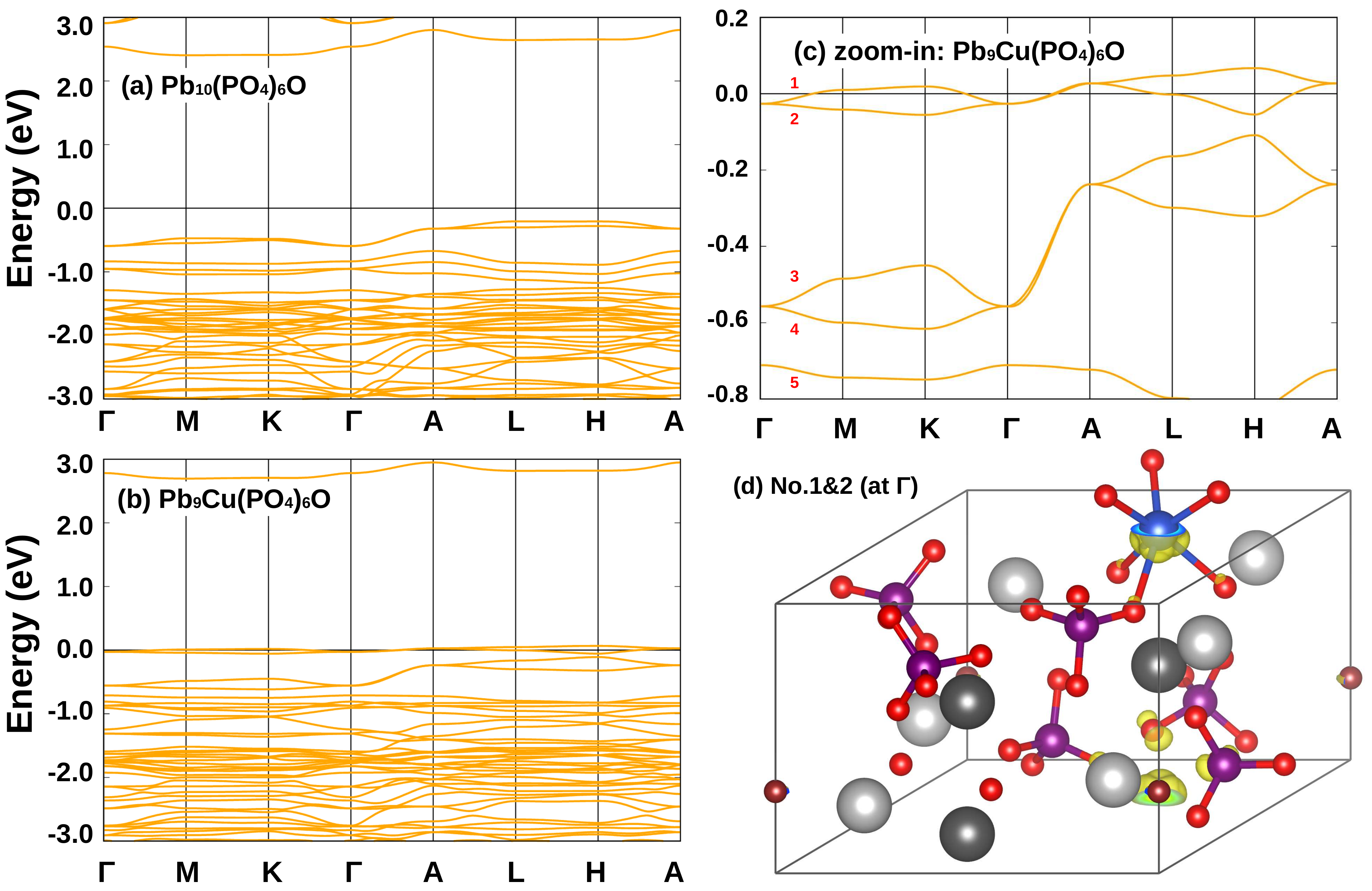}
\end{minipage}\hfill
\begin{minipage}{0.3\textwidth}
\includegraphics[width=.7\linewidth,angle=0]{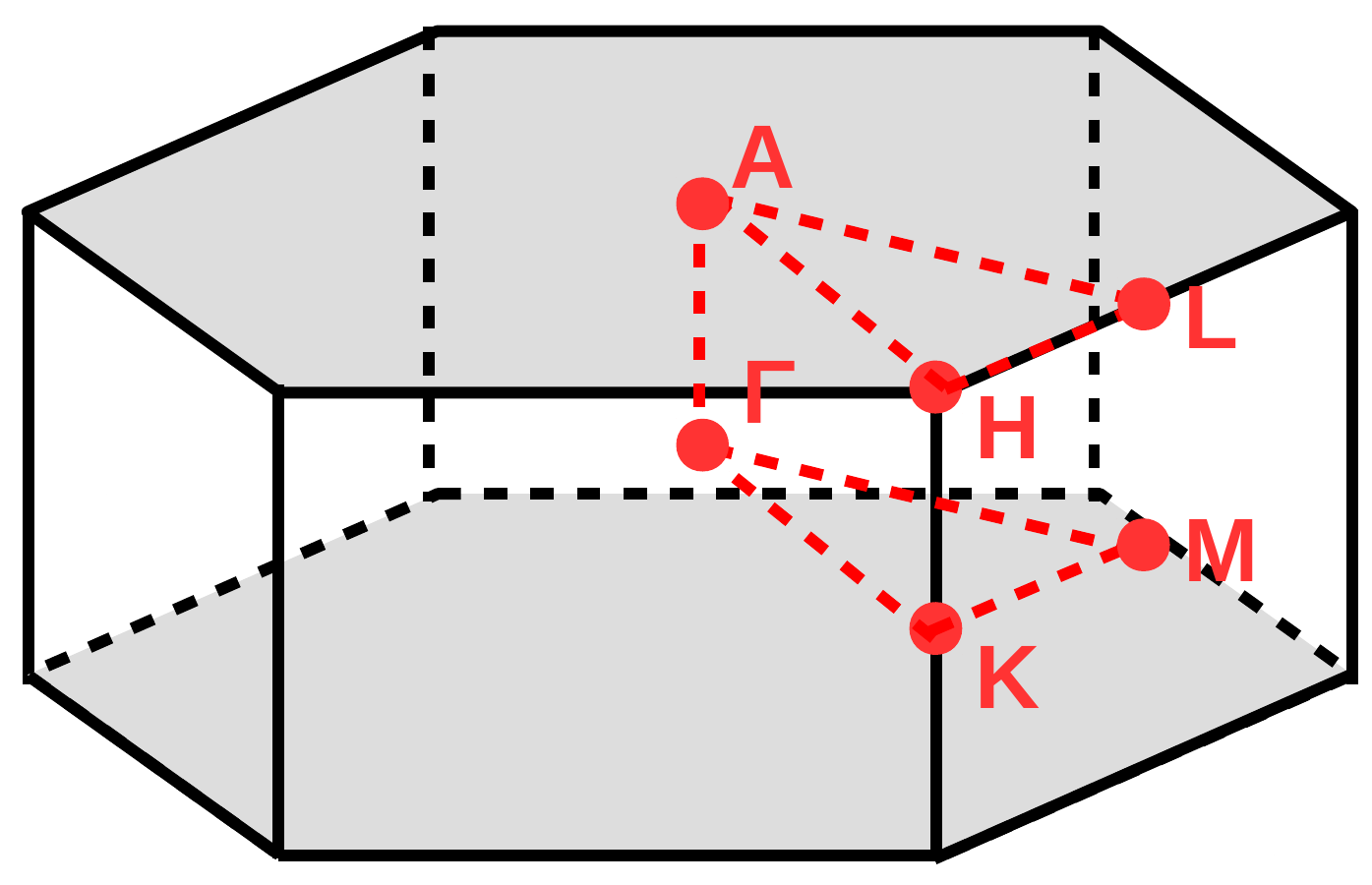}

\caption{DFT bandstructure of Pb$_{10-x}$Cu$_x$(PO$_4$)$_6$O: (a) parent compound at $x=0$; (b): putative superconductor at $x=1$); and (c): zoom-in of (b). The Fermi energy is set to zero. The top-right panel shows the high symmetry $k$-points selected for plotting the bandstructure in (a-c). Panel (d) shows the band decomposed charge density distribution (at $k$=$\Gamma$) of the two bands crossing the Fermi energy: No.~1 and No.~2 indexes the bands in (c) by their energy below the Fermi energy. 
\label{Fig:BandStructure}}
\end{minipage}
\end{figure*}

{\sl Electronic structure}---
Fig.~\ref{Fig:BandStructure} (a) shows the calculated bandstructure of the parent compound and that of the  putative superconductor.
In agreement with experiment~\cite{Lee2023_3}, the parent compound is insulating with a rather large gap of {2.3}\,eV  between the {O-$p$} and {Pb-$p$} states in DFT (see Supplementary Materials Section~II for the detailed DOS). Note, this gap may be even underestimated in DFT because for $sp$ materials the non-local exchange tends to further separate occupied from unoccupied states and thus to increase the bandgap.

When substituting one Pb by Cu, two flat bands cross the Fermi energy, labeled ``1" and ``2" in Fig.~\ref{Fig:BandStructure} (c).
The charge distribution of these flat bands in Fig.~\ref{Fig:BandStructure} (d) reveals that they originate from the Cu orbitals but with a very strong hybridization to oxygen.
These two narrow bands are occupied by 3 electrons per unit cell. Copper is thus essentially in an effective Cu$^{2+}$ state with a  3$d^9$ electronic configuration.

Due to the large Cu-Cu distance ($\approx$10\,\AA) in the lead-apatite structure, the Cu-Cu hopping is extremely small, which explains that these bands are so narrow. For example, the bandwidth of the conduction band near Fermi energy is only $\approx$120\,meV.
The small hopping certainly also contributes to the experimental observation that the high temperature metallic phase is a very bad metal with a large resistivity of $0.02 \,\Omega$cm at temperatures $T\gtrsim 380\,$K~\cite{Lee2023_1,Lee2023_3}.

\begin{figure}[tb]
\includegraphics[width=0.95\linewidth,angle=0]{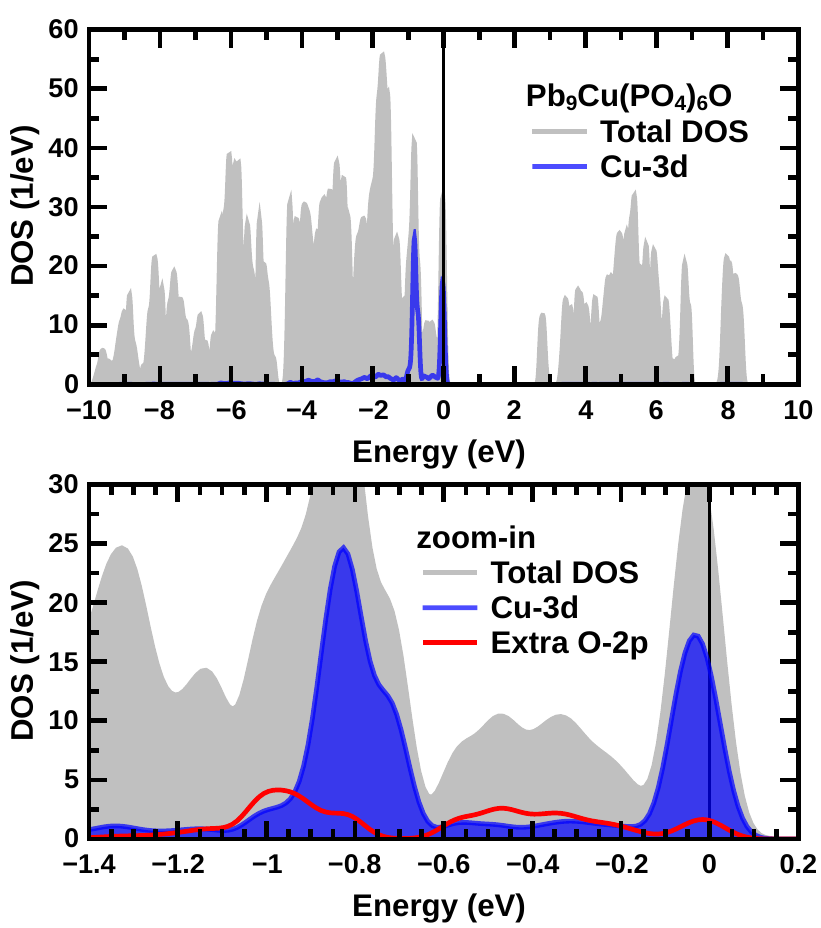}
\caption{DFT total DOS and that of Cu-3$d$ and (extra) O-2$p$ (atomically and orbitally resolved) for Pb$_{9}$Cu(PO$_4$)$_6$O. Lower panel: zoom-in. For the other partial DOSes, see Supplemental Material Section~II \cite{SM}.
\label{Fig:DOS}}
\end{figure}

Fig.~\ref{Fig:DOS} shows the total as well as the partial-Cu DOS around the Fermi energy. The other partial DOSes can be found in the Supplemental Material~\cite{SM}~Section II. The flat bands are reflected by a narrow peak of the DOS at the Fermi energy of predominately Cu-$d$ character but also with a sizable oxygen intermixing; the DOS at -0.4eV originates from the two somewhat more dispersive bands below the Fermi surface [labeled 3 and 4 in Fig.~\ref{Fig:BandStructure}~(c)] and originate predominately from the extra oxygen now with some Cu admixture, see partial DOSes and charge distribution in the Supplemental Material~\cite{SM} Section~II.

{\sl Discussion: Effects of electronic correlations}---
The low-energy electronic degrees of freedom are dominated by the two
flat Cu $d$-band crossing the Fermi energy, with a bandwidth of only 120\,meV (or hoppings $t$ of the order of 10\,meV). The local Cu $d$-$d$ interaction
will be much larger. We can expect it to be similar to that of cuprate superconductors, with a similar 3$d^9$ configuration, and $U\approx3$\,eV \cite{Yang2017,Worm2022}. If two (instead of one) low-energy Cu $d$-orbital is relevant, $U$ would even be higher because both orbitals still participates in the screening.  On the other hand, the strong oxygen intermixture to this band
can also reduce the Coulomb interaction.
In any case, this puts these two flat bands of Pb$_{9}$Cu(PO$_4$)$_6$O in an ultra-correlated regime. With $U/W\approx 25$ and for integer filling, 
the two flat bands can be expected to be split into Hubbard bands, if electronic correlations are included in e.g.~DFT+dynamical mean-field theory~\cite{Kotliar2006,held2007electronic}.
Let us emphasize that this conclusion as well
as that of a fully spin-polarized DFT+$U$ magnetic state in  Supplemental Material Section~III \cite{SM}  does not depend on the precise value of $U$; $U$ would need to be an order of magnitude smaller for obtaining a metal or a partially polarized magnetic state.

This suggests that at least a slight doping is required to arrive at a metallic system as observed in experiment \cite{Lee2023_1,Lee2023_2,Lee2023_3}.
It would put such a doped Pb$_{10-x}$Cu$_x$(PO$_4$)$_6$O compound into the category of a doped Mott or charge transfer insulator. In such a situation, we can expect a very large quasiparticle renormalization of the DFT DOS at the Fermi surface. This will further reduce the width of the flat Cu bands. Also the experimentally observed large resistivity of the metallic phase corroborates this picture.

While we can definitely say that the low-energy physics of Pb$_{10-x}$Cu$_x$(PO$_4$)$_6$O at $x\approx 1$ is dominated by predominately Cu bands, the precise arrangement of the Cu atoms and hence the lattice formed by these very Cu atoms is more difficult to predict in DFT. Periodically extending the single unit cells results in a triangular arrangement of the Cu-atoms (blue) 
in Fig.~\ref{Fig:CrystStruct}~(b). Other arrangement of the Cu atoms on the Pb(1) sites are possible and have not been studied. They will result, because of the rather larger Cu-Cu distance, in similar flat or even more flat bands. 
In the presence of disorder or for larger supercells, we would expect a similarly small bandwidth but a further suppression of the conductivity. Such a disordered arrangement of the Cu atoms is also unfavorable for long-range superconductivity.
  
The triangular lattice we find here is highly frustrated regarding magnetic fluctuations. The large distance and thus small hopping between Cu sites further suppresses the strength of prospective magnetic fluctuations. 
Also the coupling between the layers is weak: the dispersion from $\Gamma$ to $A$ is about half of that from $\Gamma$ to $M$ or $X$ in Fig.\ref{Fig:BandStructure} (c). Hence spin fluctuations cannot be expected to be particularly strong, cf.~magnetic DFT+$U$ calculations in the Supplemental Material Section~III \cite{SM}.

{\sl Discussion:  superconductivity}---
While we have not performed calculations for superconductivity, we will discuss in the following possible mechanism on the basis of the electronic structure found. That is, two flat Cu-bands crossing the Fermi energy and two somewhat more dispersive O bands immediately below.
In contrast to cuprates, the small hopping and lattice frustration suppress antiferromagnetic spin fluctuations.
This strongly speaks against spin-fluctuation as the pairing glue at elevated temperatures.  
Ferromagnetism is on the other hand known to prevail in flat bands~\cite{Mielke1993}: As $U$ dominates over $W$,  the cost in kinetic energy for a ferromagnetic arrangement becomes small compared to the energy gained by avoiding the Coulomb energy in a fully polarized ferromagnet.

Similarly and competing with ferromagnetic order, superconductivity can arise from flat bands \cite{Kuroki2005,Iglovikov2014,Aoki2020}. This flat-band mechanism is discussed, among others, as a possible mechanism for superconductivity in flat moir\'{e} bands \cite{Cao2018,Balents2020}. It is not at all clear however, whether the bandstructure of Pb$_{9}$Cu(PO$_4$)$_6$O provides for the ideal combination of flat and dispersive bands\cite{Kuroki2005,Aoki2020}. With the bands labeled 1,2 and 3,4 in Fig.~\ref{Fig:BandStructure}~(c), we have at least the necessary ingredients.

Another possibility, as already discussed in \cite{Lee2023_3} is the complicated interplay between strong electronic correlations and the BCS electron-phonon mechanism. Indeed this scenario with an increase of $T_C$ due to the enhanced quasiparticle density of states was advocated in \cite{Lee2023_3}, there coined Brinkmann-Rice-BCS mechanism \cite{Kim2021}. This scenario is somewhat hampered by the fact that also the pairing interaction gets reduced by the quasiparticle renormalization. Nonetheless, an enhancement of superconductivity by electronic correlations is possible \cite{Capone2002}. While we have not calculated the electron-phonon coupling, our DFT results attest an extraordinarily sharp peak at the Fermi energy, which is expected to be even further narrowed through such quasiparticle renormalization for the doped Mott or charge transfer insulator. Hence both of the above scenarios for superconductivity are conceivable.

Scenarios of one-dimensional superconductivity \cite{Lee2023_3} and tunneling between two-dimensional semiconductor quantum wells \cite{Lee2023_2} have also been proposed. This is not supported by the rather similar dispersion of the low-energy Cu $d$-bands in and out of plane in Fig.~\ref{Fig:BandStructure} (c).
Only the extra O bands 3,4 below the Fermi energy with a large $\Gamma$-$A$ dispersion could be considered to be one-dimensional in a first approximation.

What can we learn from our DFT calculations for  prospective non-superconducting explanations of the experimental results of  Pb$_{9}$Cu(PO$_4$)$_6$O \cite{Lee2023_1,Lee2023_2,Lee2023_3}? The sharp drop in resistivity might also occur from an ordering or structural transition, possibly affecting the lattice of the Cu dopants (at least when seeking to explain the low-$T$ resistivity of the noise level \cite{Lee2023_1,Lee2023_3}, not the $10^{-10}-10^{-11}\,\Omega$cm stated in \cite{Lee2023_2}). The putative Mei{\ss}ner effect and negative susceptibility could also result from a diamagnetic state. Here however our calculations provide some evidence against such a scenario. The narrow band(s) and the Cu 3$d^9$ electronic configuration, indicate a (only slightly) screened spin-$\frac{1}{2}$. A strong paramagnetic response can thus be expected. It is difficult to imagine how this can be overcome by a diamagnetic orbital response.

{\sl Conclusion}---
Our DFT calculations and consideration regarding correlation effects put   Pb$_9$Cu(PO$_4$)$_6$O in an ultra-strongly 
correlated regime, with  $U/W$ of ${\cal O}(10)$ instead of  ${\cal O}(1)$ in cuprate superconductors because of the very narrow Cu band(s) crossing the Fermi energy. In such a situation the Coulomb interaction $U$ clearly dominates over the kinetic energy and bandwidth $W$. This can give rise to flat-band superconductivity or a correlation-enhanced BCS mechanism. A strong diamagnetic response is, on the other hand, not expected.

It is a bit puzzling, why Pb$_{10-x}$Cu$_x$(PO$_4$)$_6$O with such a large $U/W$ was not a Mott insulator or charge transfer insulator in experiment. A possible explanation is hole or electron doping through some
off-stoichiometry (different from $x$) in experiment. Such an additional doping would put Pb$_{10-x}$Cu$_x$(PO$_4$)$_6$O into the class of doped Mott or charge transfer insulators.
Note that Pb and Cu are both $2+$; hence other dopings   $x$ will not change the Cu$^{2+}$ oxidation state; Pb$_{10-x}$Cu$_x$(PO$_4$)$_6$O remains insualting for all $x$.
This puts O (or P) deficiency or excess as a possible source for such an accidental doping off-stoichiometry. 
Given the synthesis procedure \cite{Lee2023_3} also the replacement of O or P by S is conceivable.
Against this background, it might be  advisable to actively procure such a doping in the synthesis process, e.g.\ by controlling the O partial pressure or adding small amounts of a reducing or oxidizing agent.

  {\em Note added.}
  Independently to our work, three other DFT studies appeared
  simultaneously on arXiv  \cite{lai2024first,kurleto2023pb,cabezas2023theoretical}
  but did not conclude Pb$_{9}$Cu$_1$(PO$_4$)$_6$O to be insulating.
  The insulating nature of Pb$_{10}$(PO$_4$)$_6$O, on the other hand, is
  already apparent in DFT, and has also been
  discussed in~\cite{lai2024first,yang2023ab,kurleto2023pb}.
  The Mott or charge transfer insulating state  of Pb$_{9}$Cu$_1$(PO$_4$)$_6$O has been confirmed in subsequent theoretical calculations \cite{si2023pb,korotin2023electronic,yue2023correlated,liu2023symmetry,Georgescu2023}
 and experiment \cite{kumar2023absence,puphal2023single,jiang2023pb,PhysRevMaterials.7.084804,wang2023ferromagnetic}.
 While we considered here that unintended doping leads to (super)conductivity in LK-99,  a plausible alternative  explanation is that the observed conductivity jumps are caused by remainders of Cu$_2$ in the sample \cite{Liu2023,Zhu2023,Jain2023}.

{\sl Acknowledgments}---
We acknowledge funding through the Austrian Science Funds (FWF) projects I~5398, P~36213, SFB Q-M\&S (FWF project ID F86), and Research Unit QUAST by the Deuschte Foschungsgemeinschaft (DFG; project ID FOR5249) and FWF (project ID I~5868).
L.~S.~is thankful for the starting funds from Northwest University. 
Calculations have been done on the Vienna Scientific Cluster (VSC) and super-computing clusters at Northwest University.

\onecolumngrid
\subsection{Supplemental Material to ``Electronic structure of the putative room-temperature superconductor Pb$_9$Cu(PO$_4$)$_6$O"}
\onecolumngrid
This supplementary material contains additional density-functional theory calculations regrading the lattices constants, band structure, Wannier projections and DFT+$U$ magnetic calculations. 
In Section S.I, we provide computational details and DFT structural relaxation results regrading the energetically favorable positions of extra O and doped Cu atoms. 
In Section S.II, we supplement the main text by showing all relevant partial DOS and charge distributions.
In Section S.III, we provide the DFT+$U$ total energies of different magnetic orders in Pb$_9$Cu(PO$_4$)$_6$O.

\section{Structural relaxations} 
\label{sec:details}
\subsection{Computational details}
DFT structural relaxations and electronic structure calculations are performed using the \textsc{Vasp} code \cite{PhysRevB.47.558,kresse1996efficiency} with the Perdew-Burke-Ernzerhof version for solids of the generalized gradient approximation (GGA-PBESol or GGA-PS) \cite{PhysRevLett.100.136406} for treatment for the exchange-correlation functional. 
A dense $k$-mesh of 7$\times$7$\times$9 (11$\times$11$\times$15) is employed for the unit cell structural relaxations (electronic structure) calculations.
For the calculations for supercells, the $k$-points are reduced correspondingly according to the dimensions of supercell. 
In the DFT+$U$ calculations, the interaction parameters for Cu-3$d$ orbital are $U$=7.0\,eV and $J$=0.7\,eV (adopted from Ref.~\cite{PhysRevLett.61.1415})  using the approach from Liechtenstein \emph{et al.} \cite{PhysRevB.52.R5467}. 
The plane-wave cutoff was chosen to be 500\,eV. The convergence criteria for the total energy and ionic
forces were set to 10$^{-7}$\,eV and 0.01\,eV/\AA, respectively.
Double checks for parts of the results obtained from \textsc{Vasp} are performed by using the all-electronic solution code \textsc{WIEN2K} \cite{blaha2001wien2k,Schwarz2002}. For the Wannier projections for Pb$_9$Cu(PO$_4$)$_6$O, the Cu-3$d$ bands around the Fermi energy are projected onto Wannier functions \cite{PhysRev.52.191,RevModPhys.84.1419} using \textsc{WIEN2WANNIER} \cite{mostofi2008wannier90,kunevs2010wien2wannier}.

\subsection{Position of extra O and Cu}

To investigate the ground state crystal structure of both Pb$_{10}$(PO$_4$)$_6$O and Pb$_9$Cu(PO$_4$)$_6$O, we need to answer three questions totally: (1) Has the extra O a favorable position? (2) Are
the extra O atoms in adjacent unit cells prone to occupy different, alternating positions? This is here simulated by supercell calculations. (3) Do the doped Cu atoms prone occupy different positions in a disordered or long-range pattern?

To answer the question (1), we performed a simple check: we calculate the total energy and relaxed lattice constants when the extra O occupies different positions in the units dark red or the three  white circles in Fig.~1(a) of the main text). The results indicate that the corresponding four structures are symmetrically equivalent to each other, they lead to exactly same lattice ($a$=9.825\,\AA~and $c$=7.371\,\AA) and same total energies. Indeed, this is also clear by symmetry as already mentioned in the main text. We hence conclude that for the parent compound Pb$_{10}$(PO$_4$)$_6$O, indeed the extra O in fact simply occupies one of the four possible positions, in the following parts of this study, as concluded from our DFT check, the position as labeled in Fig.~1 in main text is selected as the favorable positions for both with and without Cu-doping.

\begin{figure*}[h]
\centering
\includegraphics[width=0.9\textwidth]{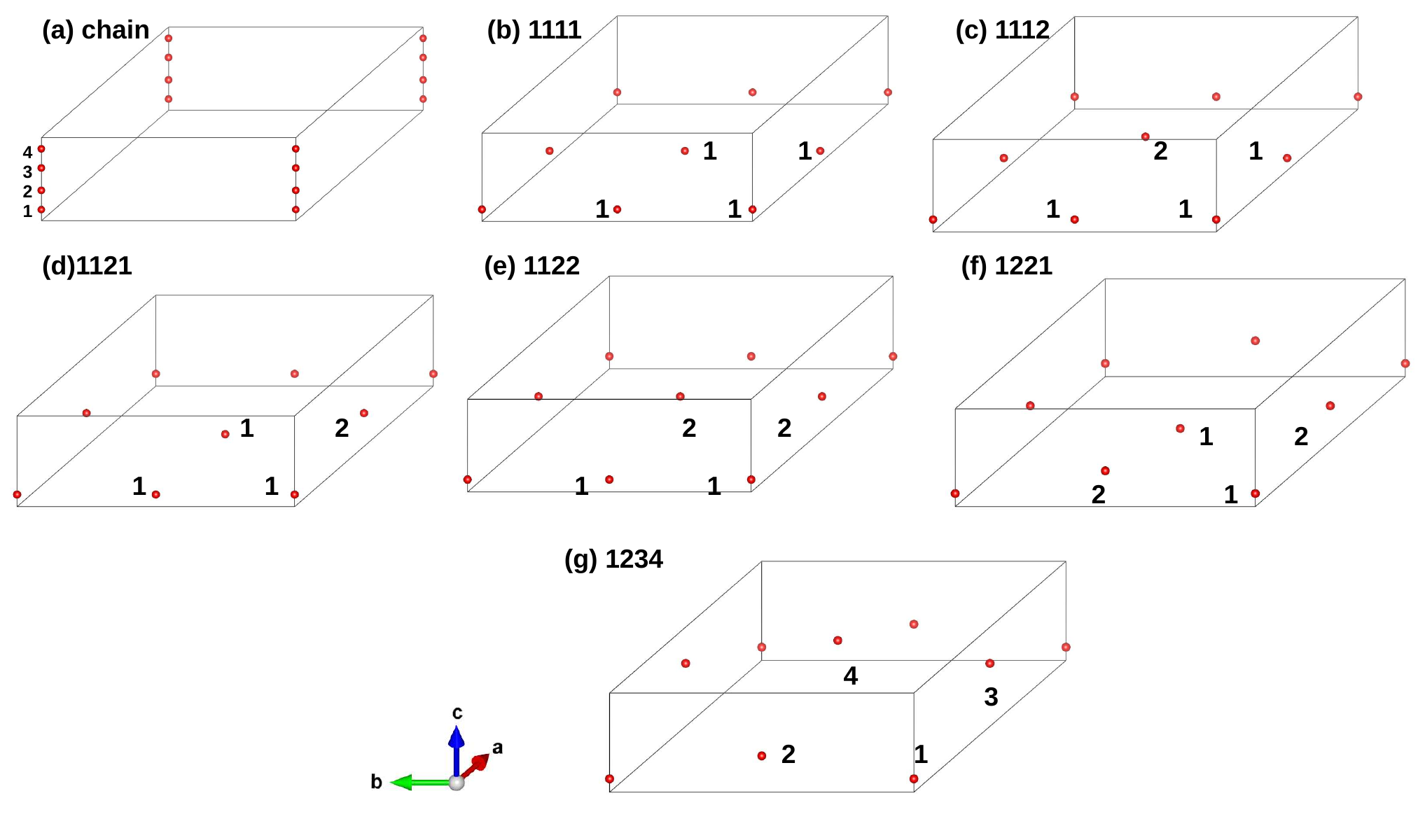}
\caption{According to the order of $z$-direction position of the extra O, we name the four possible positions in the unit cell as 1-4, as shown in (a). For a 2$\times$2$\times$1 supercell, there are four extra O atoms totally. We hence select the below 7 configurations as  representations of possible orders of the  extra Os. Please note that (a) represents a clustering behavior of extra O, i.e., the 4 extra O atoms occupy same $xy$ coordinate and form a chain along $z$ direction.}
\label{Fig_O_order}
\end{figure*}

To answer the question (2), we construct a 2$\times$2$\times$1 supercell of Pb$_{10}$(PO$_4$)$_6$O akin to that shown in Fig.~1(b) of the main text. In such a  supercell, there are 4 extra O. Considering the symmetry of Pb$_{10}$(PO$_4$)$_6$O, we selected several representations of the possible configurations of extra O atoms. We also include a clustering chain configuration as shown in Fig.~\ref{Fig_O_order}(a), where all four O atoms are in one-chain at the edge(s) of the supercell. The total energies of all possible configurations are shown in Table~\ref{Tab_O}. The ``1111" phase is by 0.051\,eV stable than the disorder phase of ``1234", while it is only $\sim$0.006\,eV (corresponding to $\sim$70\,K) higher in energy than the ``1121" phase. At room temperature and above,  temperature is higher than the latter energy and of the same order as the former. We hence conclude that he extra O in Pb$_{10}$(PO$_4$)$_6$O  will show quite some disorder distribution in its position. If not explicitly including such disorder effects, it is legitimate to restrain one-self to the case where all extra O atoms occupy the same position, as in the periodically extended unit cell in Fig.~1 (a,b) of the main text.

\begin{table*}[h]
\begin{tabular}{c|c|c|c|c|c|c|c}
\hline
\hline
Phase   & chain & 1111 & 1112 & 1121 & 1122 & 1221 & 1234 \\
\hline
Energy (eV per unit cell)    & 0.000 & -2.275 & -2.281 & -2.281 & -2.280 &  -2.281 &  -2.224 \\
\hline
\hline
\end{tabular}
\label{Tab_O}

\caption{DFT total energies of   (symmetrically inequivalent)  O configurations in the 2$\times$2$\times$1 supercell,  after structural relaxations. The energy of the ``chain" phase is set as zero.}
\end{table*}

The last question (3) is about the position of the Cu atom that substitutes one Pb atom. Based on the conclusion from Table~\ref{Tab_O} and Fig.~\ref{Fig_O_order}, we consider one unit cell which corresponds to the ``1111" oxygen configuration. For tone unit cell there are 10 possible Pb sites (some related by symmetry) the Cu atom could occupy. For these 10 possibilities we calculate the total energy of Pb$_{9}$Cu(PO$_4$)$_6$O. The computed total energies are shown in Table~\ref{Tab_Cu}. The results indicate the combination of Cu-10 and extra O at ``1" is the most energetically favorable phase. At this position (marked in blue in Fig.~1~(a) of the main text), the Cu atom is farthest away from the extra O atom.

\begin{table*}[h]
\begin{tabular}{c|c|c|c|c|c|c|c|c|c|c}
\hline
\hline
Position & Pb1 & Pb2 & Pb3 & Pb4 & Pb5 & Pb6  & Pb7 & Pb8 & Pb9 & Pb10\\
\hline
Energy (meV/unit cell) & 724.8 & 29.8 & 14.9 & 28.7 & 762.1 & 12.1  & 105.6 & 101.1 & 92.3 & 0.0 \\
\hline
\hline
\end{tabular}
\label{Tab_Cu}
\caption{DFT results of total energies for Cu occupying the 10 different Pb positions as labeled in Fig.~1~(a) of the main text. The energy of 'Pd-10' phase is set as zero.}
\end{table*}

\section{Atomically and orbital-resolved DOSes and charge distributions} 

\begin{figure}[tbh]
\includegraphics[width=.8\linewidth,angle=0]{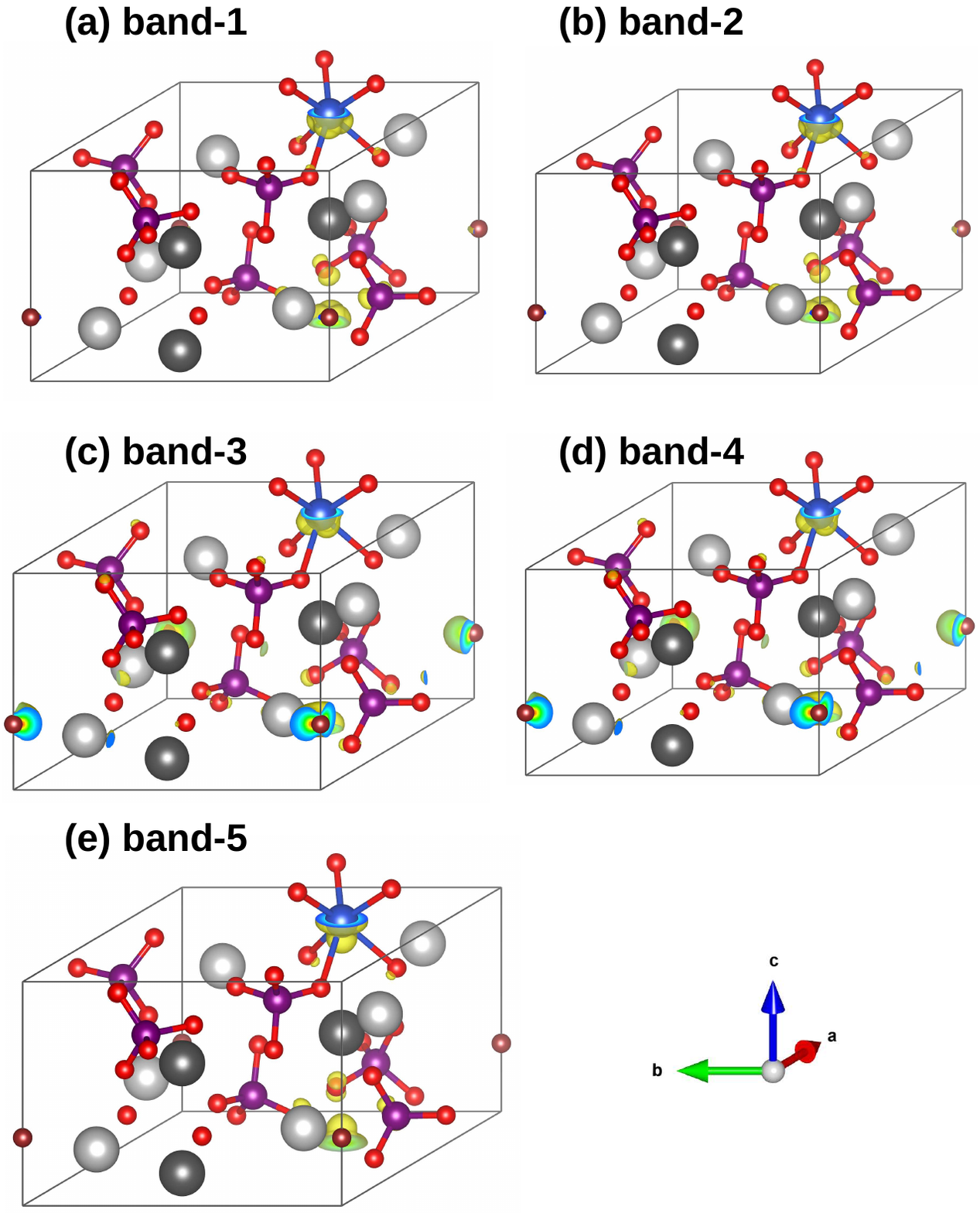}
\caption{Charge distribution within the unit cell for the $\Gamma$ state
of the 5 bands in Fig.~1~(c) of the main text crossing and below the Fermi energy.
\label{Fig:chargeSM}}
\end{figure}

Fig.~\ref{Fig:chargeSM} shows the charge distribution of the states at the $\Gamma$ point for the 5 orbitals at and below the Fermi surface that are labeled in Fig.~2(c) of the main text.
Band-1 and band-2 [that are also shown in Fig.~2(d) in main text] as well as band-3 and band-4 are degenerate at the $\Gamma$ point and thus have the same charge distribution. These four bands arise from the strong hybridization between the extra O-$p$ states and the Cu-$d$ states.

\begin{figure}[tbh]
\includegraphics[width=.85\linewidth,angle=0]{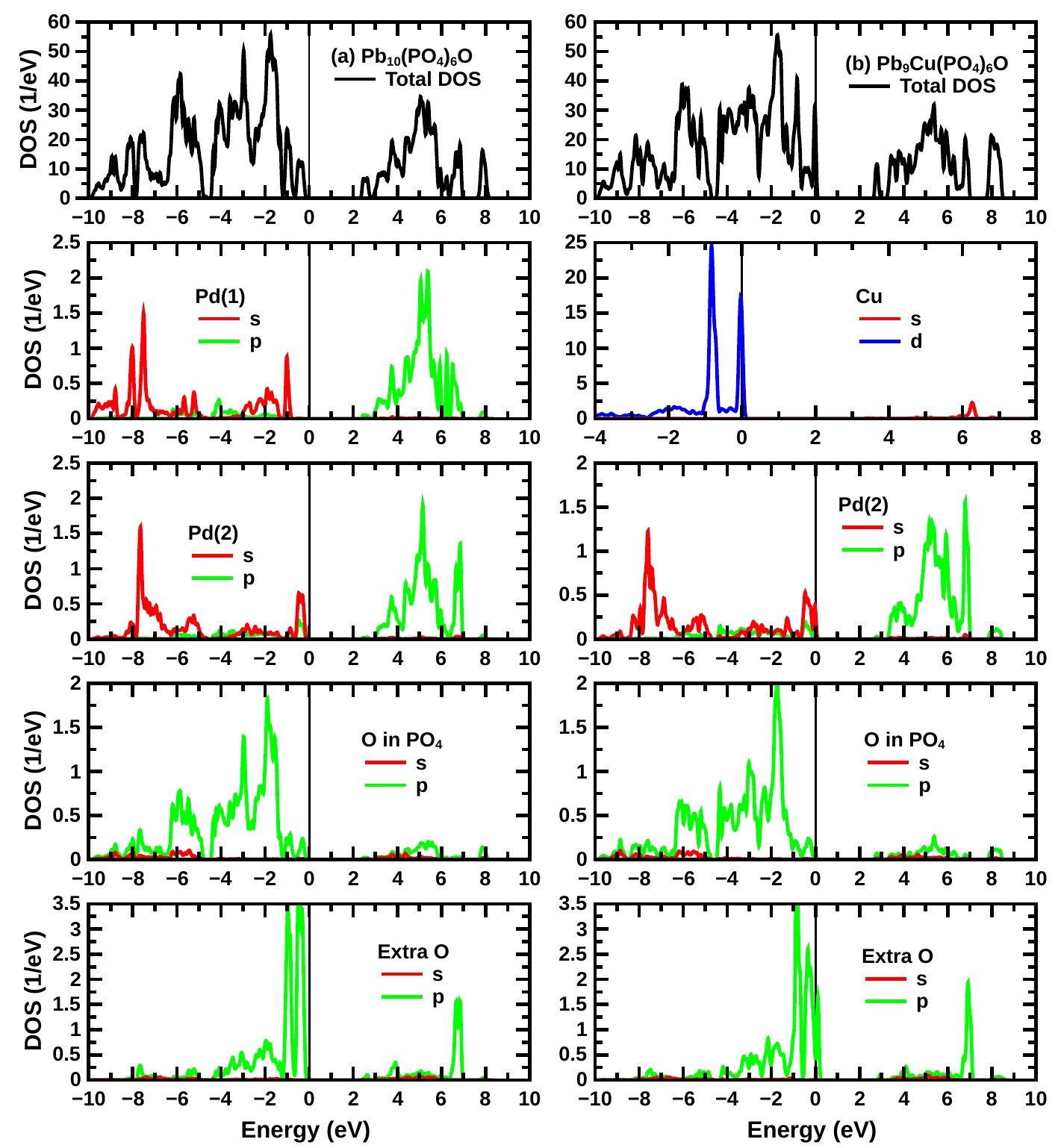}
\caption{DFT total DOS and atomically and orbitally resolved DOSes. Left:  Pb$_{10}$(PO$_4$)$_6$O; right: Pb$_{9}$Cu(PO$_4$)$_6$O. Note that the energy scale for the Cu partial DOS is enlarged and shifted.
\label{Fig:DOSSM}}
\end{figure}

Fig.~\ref{Fig:DOSSM} supplements Fig.~3 of the main text and shows the partial density of states of the other orbitals, as well as those of the parent compound.
\nopagebreak

\section{DFT+$U$ total energy and magnetic moment} 
\label{sec:dft_energy}

\begin{figure*}[h]
\centering
\includegraphics[width=0.9\textwidth]{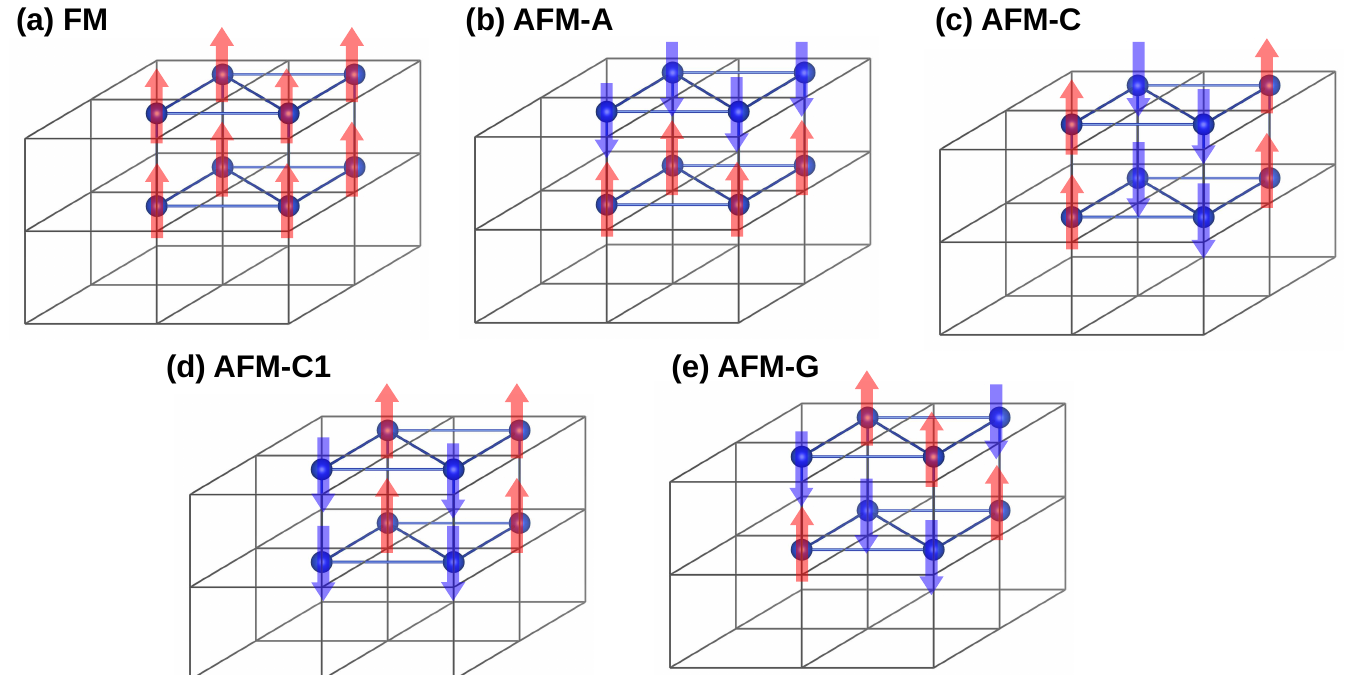}
\caption{The different magnetic orderings with moment along the  $z$-direction for which total energy calculations have been performed and are tabulated in Table~\ref{Tab1_energy}. Only the Cu sites with a corresponding moment is shown. }
\label{Fig_mag}
\end{figure*}

To check whether the spin-$\frac{1}{2}$ form stable magnetic order, we consider a 2$\times$2$\times$2 supercell that contains altogether 8 Cu ions. We then initialize different magnetic orders, including ferromagnetism (FM), and various types of antiferromagnetism (AFM): AFM-$A$, AFM-$C$, AFM-$C$1 and AFM-$G$, as shown in Fig.~\ref{Fig_mag}. With the interaction parameters of $U$=7.0\,eV and $J$=0.7\,eV, only FM and AFM-$A$ state can be converged in DFT+$U$. The total moment per supercell and the moment projected onto one Cu atom  is 8 and 0.6\,$\mu_B$ for the FM phase, and 0 and 0.6\,$\mu_B$  for the AFM-$A$, respectively. This maximal magnetic polarization on each site is a consequence of the extremely flat Cu bands at the Fermi energy. Hence fully polarizing them costs only a small amount of kinetic energy while gaining potential energy. We expect the same moments and similar total energies also for smaller $U$s.
The failures of converging other phases possibly originate from magnetic frustrations in the triangular sub-lattice of doped Cu. Indeed the triangular lattice rather suggests a 120$^{\circ}$  orientation of the AFM spins, which we have not considered here.

\begin{table*}[h]
\begin{tabular}{c|c|c|c|c|c}
\hline
\hline
Phase    & FM & AFM-$A$ & AFM-$C$ & AFM-$C$1 &  AFM-$G$  \\
\hline
Energy (meV/Cu)   &  0.0 & 3.1 & - & - &  - \\
\hline
\hline
\end{tabular}
\caption{DFT+$U$ total energies calculations and non-converged structures.}
\label{Tab1_energy}
\end{table*}

\end{document}